\documentclass{aastex6}

\shorttitle{Cosmic-ray anisotropies in right ascension}
\shortauthors{The Pierre Auger Collaboration}

\begin{document}

\title{Cosmic-ray anisotropies in right ascension \\ measured by the Pierre Auger Observatory}



\author{
A.~Aab\altaffilmark{75},
P.~Abreu\altaffilmark{67},
M.~Aglietta\altaffilmark{50,49},
I.F.M.~Albuquerque\altaffilmark{19},
J.M.~Albury\altaffilmark{12},
I.~Allekotte\altaffilmark{1},
A.~Almela\altaffilmark{8,11},
J.~Alvarez Castillo\altaffilmark{63},
J.~Alvarez-Mu\~niz\altaffilmark{74},
G.A.~Anastasi\altaffilmark{58,49}
% ...
}

\fullcollaborationName{The Pierre Auger Collaboration}

\altaffiltext{1}{Centro At\'omico Bariloche and Instituto Balseiro (CNEA-UNCuyo-CONICET), San Carlos de Bariloche, Argentina}
% ...
\altaffiltext{8}{Instituto de Tecnolog\'\i{}as en Detecci\'on y Astropart\'\i{}culas (CNEA, CONICET, UNSAM), Buenos Aires, Argentina}
% ...
\altaffiltext{11}{Universidad Tecnol\'ogica Nacional -- Facultad Regional Buenos Aires, Buenos Aires, Argentina}
\altaffiltext{12}{University of Adelaide, Adelaide, S.A., Australia}
% ...
\altaffiltext{19}{Universidade de S\~ao Paulo, Instituto de F\'\i{}sica, S\~ao Paulo, SP, Brazil}
% ...
\altaffiltext{49}{INFN, Sezione di Torino, Torino, Italy}
\altaffiltext{50}{Osservatorio Astrofisico di Torino (INAF), Torino, Italy}
% ...
\altaffiltext{58}{Universit\`a Torino, Dipartimento di Fisica, Torino, Italy}
% ...
\altaffiltext{63}{Universidad Nacional Aut\'onoma de M\'exico, M\'exico, D.F., M\'exico}
% ...
\altaffiltext{67}{Laborat\'orio de Instrumenta\c{c}\~ao e F\'\i{}sica Experimental de Part\'\i{}culas -- LIP and Instituto Superior T\'ecnico -- IST, Universidade de Lisboa -- UL, Lisboa, Portugal}
% ...
\altaffiltext{74}{Instituto Galego de F\'\i{}sica de Altas Enerx\'\i{}as (IGFAE), Universidade de Santiago de Compostela, Santiago de Compostela, Spain}
\altaffiltext{75}{IMAPP, Radboud University Nijmegen, Nijmegen, The Netherlands}
% ...

\collaboration{The Pierre Auger Collaboration}


\vfill\eject

\newpage

\begin{abstract}
We present measurements of the large-scale cosmic-ray anisotropies in right ascension, using data collected by the surface detector array of the  Pierre Auger Observatory over more than 14~years. 
We determine the equatorial dipole component, $\vec{d}_\perp$, through a Fourier analysis in right ascension that includes weights for each event so as to account for the main detector-induced systematic effects. For the energies at which the trigger efficiency of the array is small, the ``East-West'' method is employed. Besides using the data from the array with detectors separated by 1500~m, we also 
include data from the smaller but denser sub-array of detectors with 750~m separation, which allows us to extend the analysis down to $\sim 0.03$~EeV.
The most significant equatorial dipole amplitude obtained is that in the cumulative bin above 8~EeV, $d_\perp=6.0^{+1.0}_{-0.9}$\%, which is inconsistent with isotropy at the 6$\sigma$ level. 
In the bins below 8~EeV, we  obtain 99\% CL upper-bounds on $d_\perp$ at the  level of 1 to 3 percent. At energies below 1~EeV, even though the amplitudes are not significant,  the phases determined in most of the bins are not far from the right ascension of the Galactic center, at $\alpha_{\rm GC}=-94^\circ$, suggesting a predominantly Galactic origin for  anisotropies at these energies.  The reconstructed dipole phases in the energy bins above 4~EeV point instead to right ascensions that are almost opposite to the Galactic center one, indicative of an extragalactic cosmic ray origin.
\end{abstract}


\section{Introduction}
The  distribution of  cosmic-ray (CR) arrival directions is expected to provide essential clues to understanding the CR origin. Being charged particles, they are significantly deflected by the magnetic fields present in our galaxy \citep{hav15} and, for those arriving from outside it, also by the extragalactic magnetic fields \citep{fere}. Since the deflections get smaller for increasing rigidities, it is only at the highest energies that one may hope to observe localized flux excesses associated with individual CR sources. On the other hand,  as the energies lower and  the deflections become  large, the propagation  eventually becomes diffusive and it is likely that only large-scale patterns, such as a dipolar flux modulation, may be detectable. However, the small amplitudes of these anisotropies make their observation quite challenging.

Due to the Earth's rotation, cosmic-ray observatories running for long periods of time have an almost uniform exposure in right ascension. This enables them to achieve a high sensitivity to the modulation of the flux in this angular coordinate. In particular, for a dipolar cosmic-ray flux the first-harmonic modulation in right ascension provides a direct measurement of the projection of the dipole in the equatorial plane, $\vec{d}_\perp$. The possible sources of systematic uncertainties that could affect these measurements, such as those from remaining non-uniformities of the exposure or those related to the effects of atmospheric variations, can often be accounted for. Even when this is not possible, as can happen when the trigger efficiency of the array is small, methods that are insensitive to these systematic effects can be adopted to reconstruct $\vec{d}_\perp$, although they have a somewhat reduced sensitivity to the modulations. On the other hand, at low energies the  number of events detected is large, what tends to enhance the statistical sensitivity of the measurements.

The projection of the dipole along the Earth rotation axis $d_z$ can, in principle, be reconstructed from the study of the azimuthal modulation of the CR fluxes. This requires  accounting in detail for the effects of the geomagnetic field on the air showers, which can affect the reconstruction of the CR energies in an azimuthally dependent way. Also, the presence of a tilt of the array can induce a spurious contribution to $d_z$. When the trigger efficiency of the array is small, these effects may lead to systematic uncertainties that cannot be totally corrected for, particularly given the azimuthal dependence of the trigger efficiency arising from the actual geometry of the surface detector  array of the  Pierre Auger Observatory. 
Due to these limitations, we will here restrict our analysis to the  determination of  $\vec{d}_\perp$ through the study of the distribution in right ascension of the events recorded in different energy bins. We note that  
the determination of $d_z$ for energies $E\geq4$~EeV, for which that detector has full efficiency for zenith angles up to 80$^\circ$, was discussed in detail in \citet{lsa2015,science,uhedip}.

 At $E\geq8$~EeV, a significant first-harmonic modulation in right ascension, corresponding to an amplitude  $d_\perp \sim 6$\%, has been detected by the Pierre Auger Observatory \citep{science}. 
The reconstructed direction of the three-dimensional dipole suggests a predominant extragalactic origin of the CR anisotropies at  energies above 4~EeV, and the dipolar amplitudes obtained in different bins show a growing trend with increasing energies  \citep{science,uhedip}.

The phase in right ascension of the dipolar modulation of the flux determined above 8~EeV is $\alpha_d\simeq 100^\circ$. This is nearly opposite to the phases measured at PeV energies by IceCube and IceTop  \citep{ic12,ic16}, which lie not far from the Galactic center direction which is at $\alpha_{\rm GC} = -94^\circ$. Also the  KASCADE-Grande measurements, involving CR energies from few PeV up to few tens of PeV, lead to phases  lying close to the right ascension of the Galactic center, even though the measured amplitudes are not statistically significant \citep{KG}.\footnote{Hints of anisotropies on smaller angular scales were also found  recently in a reanalysis of KASCADE-Grande data \citep{ahlers}.}
All this is in agreement with the expectation that for energies above that of the knee of the CR spectrum, which corresponds to the steepening taking place at  $\sim 4$~PeV, the outward  diffusive escape of the CRs produced in the Galaxy should give rise to a dipolar flux component having its maximum not far from the Galactic center direction. Also at  energies above few EeV, where the propagation would become more rectilinear, a continuous distribution of Galactic sources should give rise to a dipolar component not far from the GC direction \citep{uhedip}. Departures from these behaviors could however result if the CR source distribution is not symmetric with respect to the Galactic center (such as in the presence of a powerful nearby CR source),  in the presence of drift motions caused by the regular Galactic magnetic field components \citep{ptus}, or when the contribution from the extragalactic component becomes sizable.  Note that the expected direction of a dipole of  extragalactic origin will depend on the (unknown)  distribution of the CR sources and on the effects of the deflections caused by the Galactic magnetic field, as was discussed in detail in \citet{uhedip}.

The change from a Galactic CR origin towards a predominantly extragalactic origin is expected to take place somewhere above the knee. More precise measurements of the large-scale anisotropies, filling the gap between the IceCube/IceTop or KASCADE-Grande measurements and the dipole determined by the Pierre Auger Observatory above 8~EeV, should provide information about this transition. In fact, although at energies below 8~EeV the reported dipolar amplitudes are not significant, indications that a change in the phase of the anisotropies in right ascension  takes place around few EeV are apparent in the Pierre Auger Observatory measurements \citep{App11,LSA2012,LSA2013,ICRC13,ICRC15}. One has to keep in mind in this discussion that
the energy at which the total anisotropy becomes of predominantly extragalactic origin may  be different from the energy at which the CR flux becomes of predominantly extragalactic origin, since the intrinsic anisotropies of each component are likely different.

We present here an  update of the measurements of the large-scale anisotropies that are sensitive to the equatorial component of a dipole, for the whole energy range from $\sim 0.03$~EeV up to $\geq32$~EeV, covering more than three decades of energy. The results above 4~EeV are an update of those presented in \citet{uhedip}, including two more years of data, corresponding to an increase in the exposure by 20\%.  At lower energies, we provide a major update of the latest published results  \citep{ICRC15}, with 50\% more exposure for the SD1500 array and twice as much for the SD750 array.   At energies below 2~EeV,  possible systematic effects related to the reduced trigger efficiency could be significant. To study the modulation in right ascension in this regime we have then to resort to the ``East-West'' method, which has larger associated uncertainties but is not affected by most of the systematic effects \citep{na89,ew}. At energies below 0.25~EeV, it proves convenient to use the data from the sub-array of detectors with 750~m spacing which, although being much smaller,  can detect a larger number of events at these energies. 

\section{The Observatory and the dataset}
The Pierre Auger Observatory \citep{NIM2015}, located near the city of Malarg\"ue in western Argentina (at latitude $35.2^\circ$ South), is the largest existing CR observatory. Its surface detector array (SD) consists of water-Cherenkov detectors having each one 12~tonnes of ultra-pure water viewed by three 9~inch phototubes. The main  array, SD1500, consists of detectors distributed  on a triangular grid with separations of 1,500~m that  span an area of 3,000~km$^2$. A smaller sub-array, SD750,  covers an area of 23~km$^2$  with detectors separated by 750~m, making it sensitive also to smaller CR energies. These arrays sample the secondary particles of the air showers reaching ground level. In addition, the fluorescence detector (FD) consists of 27 telescopes that overlook the SD array. The FD can  determine the longitudinal development of the air showers by observing the UV light emitted by atmospheric nitrogen molecules excited by the passage of the charged particles of the shower. This fluorescence light can be detected during clear moonless nights, with a corresponding duty cycle of about 15\% \citep{NIM2015}. The SD arrays have instead a continuous operation, detecting events  with a duty cycle close to 100\%. They also have a more uniform (and simpler to evaluate) exposure. This is why  the studies of the large-scale anisotropies that we perform here are based on the much larger number of events recorded by the surface arrays.

For the SD1500 array, the dataset considered in this work includes events  with energies above 0.25~EeV that were detected from 2004 January 1 up to 2018 August 31. For energies below 4~EeV, it includes events with zenith angles up to $60^\circ$, allowing coverage of 71\% of the sky, and the quality trigger applied requires that all the six detectors surrounding the one with the largest signal be active at the time the event is detected. For energies above 4~EeV,  more inclined events can be reliably reconstructed \citep{inclined} and hence the zenith-angle range is  extended up to $80^\circ$, allowing coverage of 85\% of the sky. Moreover, given that at these energies the number of detectors triggered by each shower is large (4 or more detectors for more than 99\% of the events),  we also include  in this case events passing a relaxed trigger condition, allowing that one of the six detectors that are neighbors to the one with the largest signal be missing or not functioning, provided that the reconstructed shower core be contained inside a triangle of nearby active detectors  \citep{science}. 
The integrated exposure of the array for $\theta \le 60^\circ$ and using the strict trigger selection is 60,700 km$^2$\,sr\,yr, while that for $\theta \le 80^\circ$ and relaxing the trigger is 92,500\,km$^2$\,sr\,yr. 

The CR arrival directions are reconstructed from the timing of the signals in the different triggered stations, and the angular resolution is better than 1.6$^\circ$ \citep{NIM2015}, so that it has negligible impact on the reconstruction of the dipole.
The energies of the events with $\theta\leq 60^\circ$  are assigned in terms of the reconstructed signals at a reference distance from the shower core of 1000~m. They are corrected for atmospheric effects, accounting for the pressure and air density variations following \citet{jinst17}, as well as for geomagnetic effects, following \citet{geo}. The inclined events, whose signals are dominantly produced by the muonic component of the showers, have a negligible dependence on atmospheric variations, while  geomagnetic effects are already taken into account in their reconstruction \citep{inclined}. Their energies are assigned in terms of the estimated muon number at ground level. The SD1500 array has full trigger efficiency for $E\geq 2.5$~EeV if one considers events with $\theta\leq 60^\circ$, and for $E\geq 4$~EeV for events with $\theta\leq 80^\circ$. The energies of the CRs are calibrated using the hybrid events measured simultaneously by the SD and FD detectors, in the regimes of full trigger efficiency. For lower energies, in which case we consider events with $\theta\leq 60^\circ$, the energy assignment is performed using the extrapolation of the corresponding calibration curve. The energy resolution for events  with $\theta\leq 60^\circ$ is about 7\% above 10~EeV, and degrades for lower energies, reaching about 20\% at 1~EeV, while the systematic uncertainty in the energy scale is 14\% (see  \citet{spectrum} for details). The more inclined events have an energy resolution of 19\%, with a similar systematic uncertainty \citep{inclined}.

For energies below 0.25~EeV, and down to $\sim 0.03$~EeV (below which the trigger efficiency is tiny), we use the events from the denser and smaller SD750 array, since the accumulated statistics is larger. The dataset comprises events  with zenith angles up to $55^\circ$ detected from 2012 January~1 up to 2018 August~31. The trigger applied requires that all six detectors  around the one with the largest signal be functioning and the associated exposure  is 234~km$^2$\,sr\,yr. The energies are assigned in terms of the reconstructed signals at a reference distance from the shower core of 450~m. They are corrected for atmospheric  effects following \citet{jinst17}. The SD750 array has full trigger efficiency for $E\geq 0.3$~EeV if one considers events with $\theta\leq 55^\circ$ \citep{NIM2015}. The energies are calibrated with hybrid events observed in the regime of full trigger efficiency and below that threshold the energy assignment is performed on the basis of the extrapolation of the corresponding calibration curve. At 0.3~EeV the energy resolution is about 18\% \citep{coleman}.

\section{The analysis method}
The weighted first-harmonic  analysis in the right ascension angle $\alpha$, often referred to as Rayleigh analysis, provides 
the Fourier coefficients as
\begin{equation}\label{eq:fcoef}
a=\frac{2}{\mathcal{N}}\sum_{i=1}^N w_i\cos \alpha_i ,~~~~~~
b=\frac{2}{\mathcal{N}}\sum_{i=1}^N w_i\sin \alpha_i,
\end{equation}
where the sums run over all $N$ detected events. The weights $w_i$, which are of order unity, account for the effects of the non-uniformities in the exposure as a function of time, with the normalization factor being ${\mathcal{N}}\equiv \sum_i w_i$. The amplitude and phase of the first-harmonic modulation are given by $r=\sqrt{a^2+b^2}$ and $\varphi=\arctan(b/a)$. The probability to obtain an amplitude larger than the one measured  as a result of a fluctuation from an isotropic distribution is $P(\ge r)=\exp(-\mathcal{N}r^2/4)$. 
To obtain the weights, we permanently monitor the number of active unitary detector cells, corresponding to the number of active detectors that are surrounded by an hexagon of working detectors or, when considering the relaxed trigger condition above 4~EeV, we also account for detector configurations with only five active detectors around the central one.
We obtain from this the exposure of the Auger Observatory in bins of right ascension of the zenith of the array, $\alpha^0$. This angle is given by $\alpha^0(t_i)\equiv 2\pi t_i/T_s ~ (\mathrm{mod} ~ 2\pi)$, with the origin of time being taken such that $\alpha^0(0)=0$. The sidereal-time period, $T_s\simeq 23.934$~h, corresponds to one extra cycle per year with respect to the solar frequency. The  fraction of the total exposure that is associated to a given $\alpha^0$ bin, taken to be of 1.25$^\circ$  width (5 minutes), is proportional to
 the total number of unitary cells in that bin,  $N_{\mathrm{cell}}(\alpha^0)$. 
The weights $w_i$  account for the relative variations of $N_{\mathrm{cell}}$ as a function of $\alpha^0$, i.e. 
\begin{equation}
w_i
=\left(\frac{N_{\mathrm{cell}}(\alpha^0(t_i))}{\langle N_{\mathrm{cell}}\rangle}\right)^{-1},
\end{equation}
with $\langle N_{\mathrm{cell}}\rangle=1/(2\pi)\int_0^{2\pi} \mathrm{d}\alpha^0~ N_{\mathrm{cell}}(\alpha^0)$.  Including these weights in the Fourier coefficients eliminates the spurious contribution to the amplitudes associated to the non-uniform exposure in right ascension.

We note that if one were to consider periods of only a few months, the  resulting modulation  of  $N_{\mathrm{cell}}(\alpha^0)$  could amount to an effect of a few percent on the modulation in right ascension of the event rates. However, after considering several years, the modulations that appear on shorter time scales tend to get averaged out, with the surviving effects being now typically at the level of about $\pm 0.5$\%.
The effects of the tilt of the SD array \citep{LSA2012}, which is inclined on average by $\sim 0.2^\circ$ towards $\phi\simeq-30^\circ$ (i.e. towards the South-East), can also be accounted for by adding an extra factor in  the weights \citep{uhedip}. However, this is actually only relevant when performing the Fourier analysis in the azimuth variable $\phi$, something we will not perform here.

When the triggering of the array is not fully efficient, there are additional systematic effects related to the interplay between the atmospheric effects in the air-shower development and the energy-dependent trigger efficiency. In particular, changes in the air density modify the Moli\`ere radius determining the lateral spread of the electromagnetic component of the showers.
The fall-off of the signal at ground level is preferentially harder under hot weather conditions and steeper under cold ones. The detection efficiency of the SD is thus expected to follow these variations to some extent,
being on average larger  when the weather is hot than when it is cold. As a consequence, one could expect that, at energies below full trigger efficiency, a spurious modulation could appear at the solar frequency.

Moreover, we have found that the amplitude of the modulation of the rates at the antisidereal frequency, which is that corresponding to one cycle less per year than the solar frequency, suggests that  spurious unaccounted effects  become relevant below 2~EeV. In particular, the Fourier amplitude corresponding to the antisidereal time period $T_{\rm as}=24.066$~h  in the bin [1, 2]~EeV is $r=0.005$. This has a probability of arising as a fluctuation of less than 0.1\%. 
A non-negligible antisidereal amplitude could for instance appear in the presence of daily and seasonal systematic effects which are not totally accounted for.
Since in this case comparable spurious amplitudes could be expected in the sidereal and antisidereal sidebands  \citep{fast}, we only use the Rayleigh method described before in the bins above 2~EeV. We have checked that in the bins above 2~EeV the amplitudes at both the solar and antisidereal frequencies are consistent with being just due to fluctuations, so that there are no signs indicating that surviving systematic effects could be present at the sidereal frequency at these energies (see Table~\ref{tab:sol-asid} in the Appendix).\footnote{Given that, for events with zenith angles smaller than 60$^\circ$, the trigger efficiency is larger than $\sim 95$\% above 2~EeV, the efficiency related systematic effects are negligible above this threshold.}
Alternatively, one can use for the energies below 2~EeV the differential {East-West} (EW) method \citep{ew}, which  is based on the difference between the counting rates of the events measured from the East sector and those from the West sector. Since the exposure is the same for events coming from the East and for those coming from the West\footnote{A possible tilt of the array in the East-West direction, giving just a constant term in the East-West rate difference, does not affect the determination of the first-harmonic modulation.}, and also the spurious modulations due to the atmospheric effects are the same in both sectors, the relative difference between both rates, $(E-W)/(E+W)$,  is not sensitive to these experimental and atmospheric systematic effects. This allows one to reconstruct in a clean way the modulation of the rate itself, without the need to apply any correction but at the expense of a reduced sensitivity to the amplitude of the CR flux  modulations. 

In this approach  \citep{ew}, the first-harmonic amplitude and phase are calculated using a slightly modified Fourier analysis  that  accounts for the subtraction of the Western sector from the Eastern one. The Fourier coefficients are defined as
\begin{equation}
a_{\rm EW}=\frac{2}{N}\sum_{i=1}^N \cos(\alpha^0(t_i)-\xi_i),~~~~~~b_{\rm EW}=\frac{2}{N}\sum_{i=1}^N \sin(\alpha^0(t_i)-\xi_i),
\end{equation}
where  
$\xi_i=0$ for events coming from the East and $\xi_i=\pi$ for those coming from the West, so as to easily implement the subtraction of data from the two hemispheres.  

In the case in which the dominant contribution to the flux modulation is purely dipolar, the amplitude $r_{\rm EW}=\sqrt{a_{\rm EW}^2+b_{\rm EW}^2}$ and phase $\varphi_{\rm EW}=\arctan(b_{\rm EW}/a_{\rm EW})$ obtained with this method are related to the ones from the Rayleigh formalism through $r=\frac{\pi\langle\cos\delta\rangle}{2\langle\sin\theta\rangle}r_{\rm EW}$ and $\varphi=\varphi_{\rm EW}+\pi/2$, where $\langle\cos\delta\rangle$ is the average of the cosine of the declination of the events and similarly $\langle\sin\theta\rangle$ is the average of the sine of their zenith angles   \citep{ew}. The probability to obtain an amplitude larger than the one measured  as a result of a fluctuation from an isotropic distribution is $P(\ge r_{\rm EW})=\exp(-Nr_{\rm EW}^2/4)$.

The amplitude of the equatorial dipole component is related to the amplitude of the first-harmonic modulation through  $d_\perp\simeq r/\langle\cos\delta\rangle$, and its phase $\alpha_d$  coincides with the first-harmonic phase $\varphi$.

\section{Right ascension modulation from 0.03~E\MakeLowercase{e}V up to $E\geq 32$~E\MakeLowercase{e}V}

In Table~\ref{tab:dper}, we report the results for the reconstructed equatorial dipole in different energy bins, covering the range from $\sim  0.03$~EeV up to $E\geq 32$~EeV. The energies defining the boundaries of the bins are $2^n$\,EeV, with $n=-5,-4,...,4,5$. As mentioned previously, the results are obtained from the study of the right ascension modulation using different methods and datasets. We use the weighted Rayleigh analysis  in the energy bins above 2~EeV, for which the systematic effects associated with the non-saturated detector efficiency and to the effects related to  atmospheric variations are well under control. When this is not the case, we report the results of the East-West method which, although having larger uncertainties, is quite insensitive to most sources of systematic effects in  the right ascension distribution. For energies above 0.25~EeV, we report the results obtained with the data from the SD1500 array, while for lower energies we use the dataset from the SD750 array since, having a lower threshold, it leads to a larger number of events despite the reduced size of the array. In that case, given that the SD750 array is not fully efficient below 0.3~EeV,  we just use the East-West method. 

\begin{table}[h]
\centering
\resizebox{\textwidth}{!}{
\begin{tabular}{c | c c c c c c c c}
  & $ E$ [EeV] & $ E_{\rm med}$ [EeV] & $N$ &  $d_\perp$ [\%] & $\sigma_{x,y}$ [\%] & $\alpha_d [^\circ]$ & $P(\ge d_\perp)$ & 
  $d_\perp^{\rm UL}$ [\%] \\
\hline
\rule{0pt}{3ex} 
East-West & 1/32 - 1/16 & 0.051 & 432,155 & $1.0^{+1.0}_{-0.4}$ & 0.91 & $112 \pm 71$ & 0.54 &
3.3 \\
(SD750) & 1/16 - 1/8 & 0.088 & 924,856 & $0.6^{+0.6}_{-0.3}$ & 0.52& $-44 \pm 68$ & 0.50 
& 2.0 \\
 & 1/8 - 1/4  & 0.161 & 488,752  & 
  $ 0.2^{+0.8}_{-0.2}$& 0.63 & $-31 \pm 108$ & 0.94 & 
  2.0 \\  
\hline
\rule{0pt}{3ex} 
East-West &1/4 - 1/2  & 0.43 & 770,316 & 
$0.6^{+0.5}_{-0.3}$ & 0.48 & $-135 \pm 64$ & 0.45 &
1.8 \\ 
 (SD1500) & 1/2 - 1  & 0.70 & 2,388,467 & $0.5^{+0.3}_{-0.2}$ & 0.27& $-99 \pm 43$ & 0.20 &  
 1.1 \\
& 1 - 2 & 1.28 & 1,243,103 & $0.18^{+0.47}_{-0.02}$ & 0.35& $-69 \pm 100$ & 0.87 &
1.1 \\
\hline
\rule{0pt}{3ex} 
Rayleigh & 2 - 4 & 2.48 & 283,074 & $0.5^{+0.4}_{-0.2}$ & 0.34 & $-11 \pm 55$ & 0.34 &
1.4 \\
(SD1500) & 4 - 8  & 5.1 & 88,325 & $1.0^{+0.7}_{-0.4}$ &0.61 & $69 \pm 46$ & 0.23 &
2.6  \\
 &  8 - 16  & 10.3 & 27,271 & $5.6^{+1.2}_{-1.0}$ &1.1 & $97 \pm 12$ & $2.3\times 10^{-6}$ &
  -- \\ 
  & 16 - 32  & 20.3 & 7,664 & $7.5^{+2.3}_{-1.8}$ & 2.1& $80 \pm 17$ & $1.5\times 10^{-3}$ &
  -- \\ 
   & $\geq 32$  & 40 &  1,993 & $13^{+5}_{-3}$ & 4.1 & $152 \pm 19$ & $5.3\times 10^{-3}$ &
    -- \\ 
   \cline{2-9}
  \rule{0pt}{3ex} 
 
  & $\geq 8$ & 11.5 & 36,928 & $6.0^{+1.0}_{-0.9}$ & 0.94& $98 \pm 9$ & $1.4\times 10^{-9}$ & 
 -- \\
\end{tabular}
}
\caption{Equatorial dipole reconstruction in different energy bins.  Indicated are the median energies in each bin $E_{\rm med}$, number of events $N$, amplitude of $d_\perp$, uncertainty $\sigma_{x,y}=\sigma/\langle\cos\delta\rangle$ of the components $d_x$ or $d_y$, right ascension phase, probability to get a larger amplitude from fluctuations of an isotropic distribution and 99\% CL upper limit on the amplitude.}
\label{tab:dper}
\end{table}

\begin{figure}[t]
\centering
\includegraphics[scale=1.03]{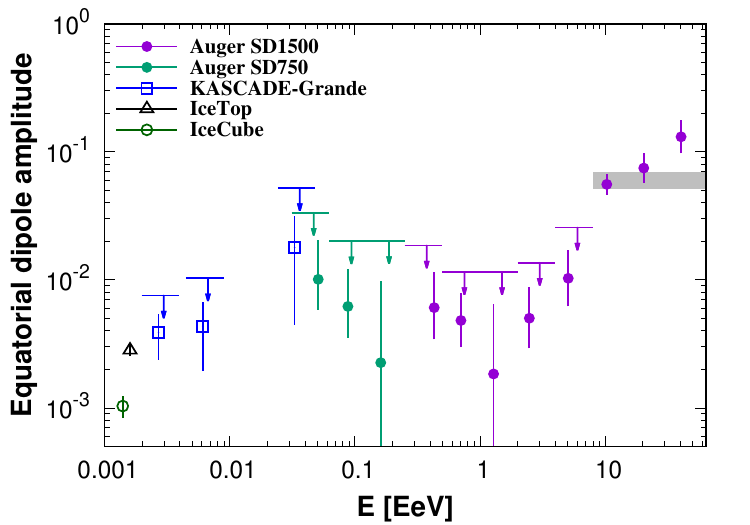}
\includegraphics[scale=1.03]{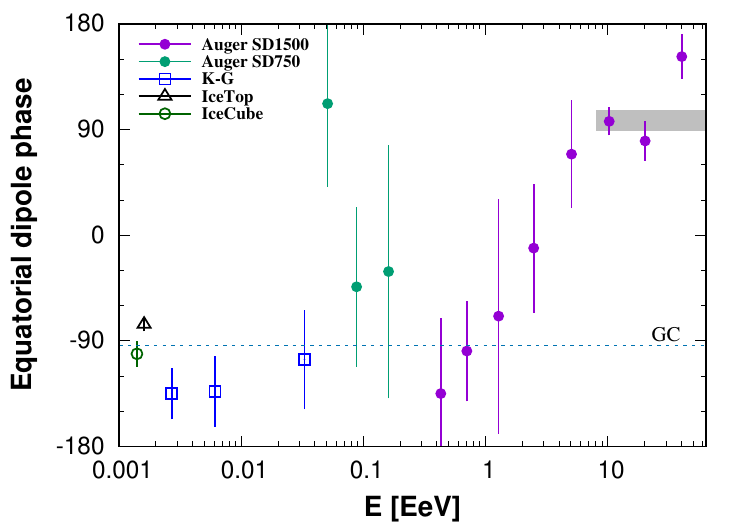}
\caption{Reconstructed equatorial-dipole amplitude (left) and phase (right). The upper limits at 99\%~CL are shown for all the energy bins in which the measured amplitude has a chance  probability greater than 1\%. The gray bands indicate the amplitude and phase for the energy bin $E\geq 8$~EeV. Results from other experiments are shown for comparison \citep{ic12,ic16,KG}.}
\label{fig:dper}
\end{figure}

For each energy bin, we report in  Table~\ref{tab:dper} the number of events $N$, the amplitude $d_\perp$, the  uncertainty $\sigma_{x,y}$ of the components $d_x$ or $d_y$, the right ascension phase of the dipolar modulation $\alpha_d$, the chance probability $P( \ge d_\perp)$ and, when the measured amplitude has a probability larger than 1\%, we also report the 99\% CL upper limit   on the amplitude of the equatorial dipole  $d_\perp^{\rm\small UL}$. 
The upper limits on the first-harmonic amplitude at a given confidence level CL (${\rm CL}=0.99$ for 99\% CL) are derived from the distribution for a dipolar anisotropy of unknown amplitude,  marginalized over the dipole phase, requiring that
\begin{equation}
\int_{0}^{r^{\rm\small UL}}\mathrm{d}r\,\frac{r}{\sigma^2}\exp\left[-\frac{r^2+s^2}{2\sigma^2}\right]I_0\left(\frac{rs}{\sigma^2}\right) = {\rm CL},
\end{equation}
with $I_0(x)$ the zero-order modified Bessel function, $s$ the measured amplitude and  the dispersion being $\sigma=\sqrt{2/\mathcal{N}}$ for the Rayleigh analysis while $\sigma=(\pi\langle\cos\delta\rangle/2\langle\sin\theta\rangle)\sqrt{2/N}$ for the East-West method. These bounds on the first-harmonic amplitude are then converted into the corresponding upper limit for the amplitude of the equatorial dipole using that $d_\perp^{\rm\small UL}=r^{\rm\small UL}/\langle\cos\delta\rangle$. For the uncertainties in the phase,  we use the two-dimensional distribution marginalized instead over the dipole amplitude $r$ \citep{linsley}. In Table~\ref{tab:dperew} in the Appendix we also report the results obtained above 2~EeV with the East-West method, which are consistent with those obtained with the Fourier analysis in  Table~\ref{tab:dper} but have larger uncertainties. 

Fig.~\ref{fig:dper} shows  the equatorial dipole amplitude (left panel) and phase (right panel) that were determined in all the energy bins considered, as reported in Table~\ref{tab:dper}. Also shown are the results obtained by the IceCube, IceTop and KASCADE-Grande experiments in the 1--30~PeV range \citep{ic12,ic16,KG}.  We also show the 99\% CL upper limit $d_\perp^{\rm UL}$ in the cases in which the measured amplitude has more than 1\% probability to be a fluctuation from an isotropic distribution. The results for the integral bin with $E\geq8$~EeV, that was considered in \citet{science}, is shown as a gray band.

A trend of increasing amplitudes for increasing energies is observed, with values going from $d_\perp\simeq 0.1$\% at PeV energies, to $\sim 1$\% at EeV energies and reaching $\sim 10$\% at 30~EeV. Regarding the phases, a transition between values lying  close to the right ascension of the Galactic center,  $\alpha_d\simeq \alpha_{\rm GC}$,  towards values in a nearly opposite direction, $\alpha_d\simeq 100^\circ$, is observed to take place around a few EeV. 

\begin{figure}[h]
\centering
\includegraphics[scale=.44]{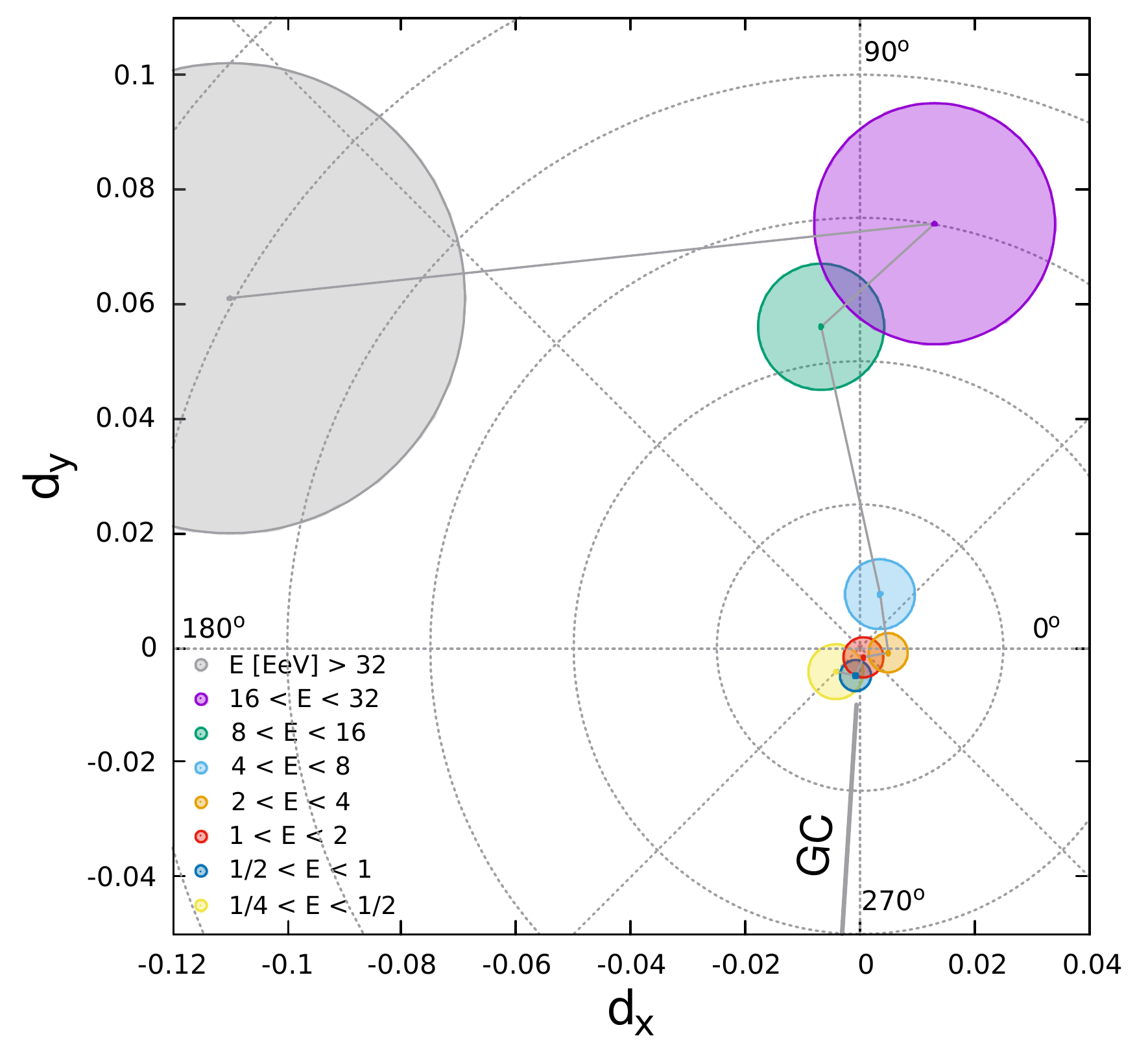}
\includegraphics[scale=.44]{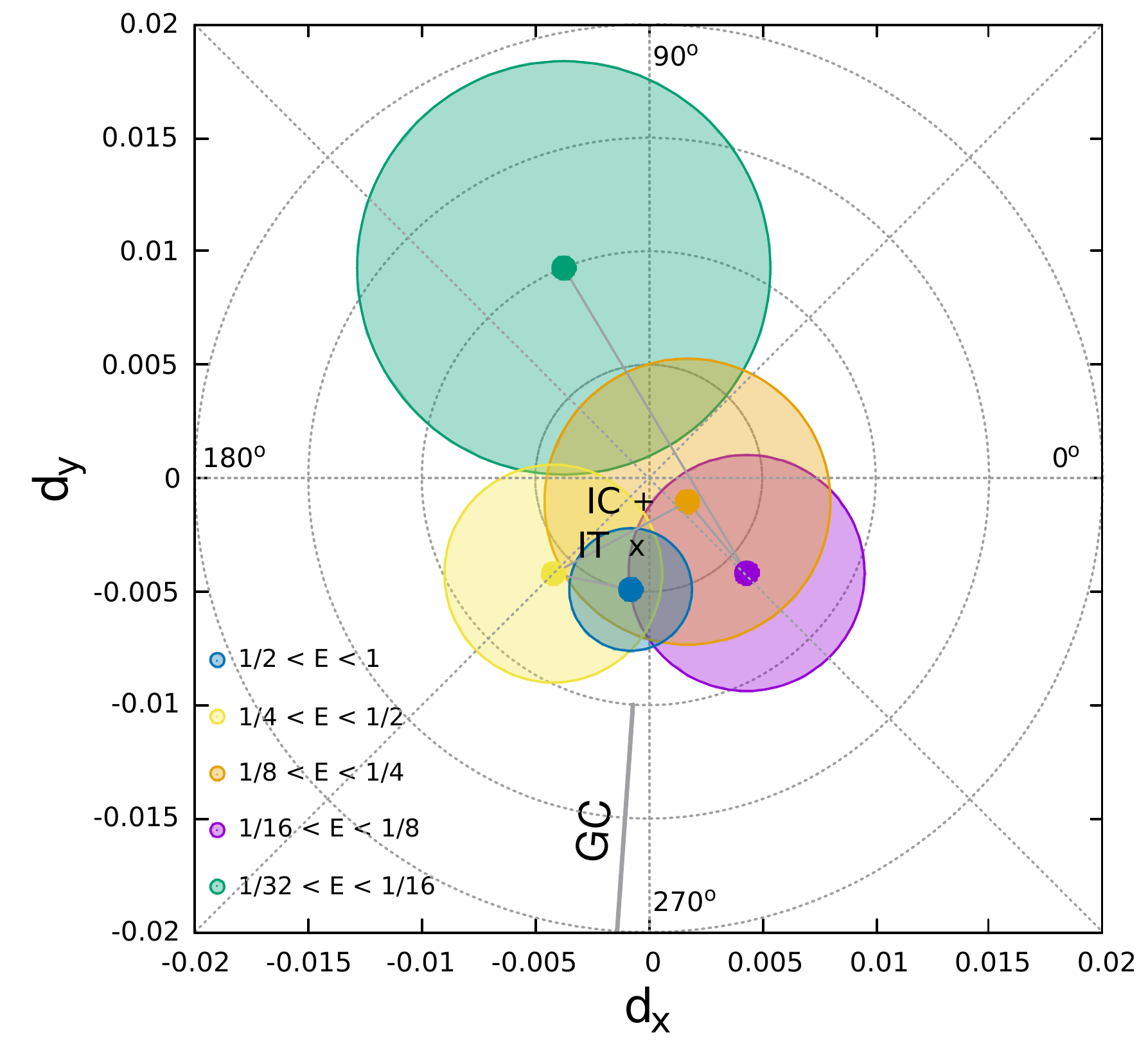}

\caption{Components of the dipole in the equatorial plane for different energy bins above 0.25~EeV (left panel) and below 1~EeV (right panel). The horizontal axis corresponds to the component along the direction $\alpha=0$ while the vertical axis to that along $\alpha=90^\circ$.  The radius of each circle corresponds to the 1$\sigma$ uncertainty in $d_x$ and $d_y$. The Galactic center direction is also indicated. The measurements from IceCube (IC) and IceTop (IT) at PeV energies are also indicated in the right panel \citep{ic12,ic16}.}
\label{fig:circles}
\end{figure}

The overall behavior of the amplitudes and phases  in the $d_x$--$d_y$  plane is depicted in Fig.~\ref{fig:circles}. The left panel includes the  energies above 0.25~EeV while the right panel those below 1~EeV. In these plots, the right ascension $\alpha_d$ is the polar angle, measured anti-clockwise from the x-axis (so that $d_x=d_\perp \cos\alpha_d$ and $d_y=d_\perp\sin\alpha_d$). The circles shown  have a radius equal to 
the 1$\sigma$ uncertainties $\sigma_{x,y}$ in the dipole components $d_{x,y}$ (reported in the Table~1), effectively including $\sim 39$\% of the two-dimensional confidence region. One can appreciate in this plot how the amplitudes decrease for decreasing energies, and how the phases change as a function of the energy, pointing almost in the opposite direction of the Galactic center above 4~EeV and not far from it below 1~EeV.

The values of the anisotropy parameters obtained above are based, by construction, on the event content in the energy intervals under scrutiny. The finite resolution on the energies induces bin-to-bin migration of events. Due to the steepness of the energy spectrum, the migration happens especially from lower to higher energy bins. This influences the energy dependence of the recovered parameters. However, given that the size of the energy bins chosen here is much larger than the resolution, the migration of events remains small enough to avoid significant distortions for the recovered values above full efficiency. For instance, given the energy resolution of the SD1500 array \citep{spectrum} and assuming a dipole amplitude scaling as $E^{0.8}$, as was found in \citet{uhedip}  to approximately hold above 4~EeV, the impact of the migrations remains below an order of magnitude smaller than the statistical uncertainties associated to the recovered parameters. In the energy range below full efficiency, additional systematic effects enter into play on the energy estimate. We note that forward-folding simulations of the response function effects into an injected anisotropy show that the recovered parameters are not impacted by more than their current statistical uncertainties. A complete unfolding of these effects is left for future studies. It requires an accurate knowledge of the response function of the SD arrays down to low energies, which is not available at the moment.

\section{Discussion and conclusions}
We have updated the searches for anisotropies on large angular scales using the cosmic rays detected by the Pierre Auger Observatory. The analysis covered more than three orders of magnitude in energy, including events with  $E\geq 0.03$~EeV and hence
 encompassing the expected transition between Galactic and  extragalactic origins of the cosmic rays. This was achieved by studying the first-harmonic modulation in right ascension of the CR fluxes determined with the SD1500 and the SD750 surface detector arrays. This allowed us to determine the equatorial component of a dipolar modulation, $\vec{d}_\perp$, or eventually to set strict upper-bounds on it.

For the inclusive bin above 8~EeV, the first-harmonic modulation in right ascension leads to an equatorial dipole amplitude $d_\perp= 0.060^{+0.010}_{-0.009}$, which has a  probability to arise by chance from an isotropic distribution of $1.4\times 10^{-9}$, corresponding to a two-sided Gaussian significance of 6$\sigma$. The phase of the maximum of this modulation is at $\alpha_d= 98^\circ\pm 9^\circ$, indicating  an extragalactic origin for these CRs.
 When splitting the bin above 8~EeV, as originally done in  \citet{uhedip}, one finds indications of an increasing amplitude with increasing energies, and the direction of the dipole suggests that it has an extragalactic origin in all the three bins considered. A  growing  dipole amplitude 
 for increasing energies could for instance be associated  with the larger relative contribution to the flux that arises at high energies from nearby sources, that are more anisotropically distributed than the integrated flux from the distant ones. A suppression of the more isotropic contribution from distant sources is expected to result from the strong attenuation of the CR flux that should take place at the highest energies as a consequence of their interactions with the  background radiation \citep{gzk1,gzk2}.

At energies below 8~EeV, none of the amplitudes are significant, and we set 99\% CL upper bounds on $d_\perp$ at the level of 1 to 3\%. The phases measured in most of the bins below 1~EeV  are not far from the direction towards the Galactic center.
All this suggests that the origin of these dipolar anisotropies changes from a predominantly Galactic one to an extragalactic one somewhere in the range between 1~EeV and few~EeV. 
The small size of the dipolar amplitudes in this energy range, combined with the indications that the composition is relatively light \citep{compo}, disfavor a predominant flux component of Galactic origin  at $E>1$~EeV \citep{LSA2013}. 
Models of Galactic CRs relying on a mixed mass composition, with rigidity dependent spectra, have been proposed to explain the knee (at $\sim 4$~PeV) and second-knee (at $\sim 0.1$~EeV) features in the spectrum \citep{candia}. The predicted anisotropies depend on the details of the Galactic magnetic field model considered and, below 0.5~EeV, they are consistent with the upper bounds we obtained. An extrapolation of these models, considering that there is no cutoff in the Galactic component, would predict dipolar anisotropies at the several percent level  beyond the EeV, in tension with the upper bounds in this range. The conflict is even stronger for Galactic models  \citep{kusenko} having a light CR composition that extends up to the ankle energy (at $\sim 5$~EeV). The presence of a  more isotropic extragalactic component making a significant contribution  already at EeV energies could dilute the anisotropy of Galactic origin, so as to be consistent with the bounds obtained. Note that even if the extragalactic component were completely isotropic in some reference frame,  the motion of the Earth with respect to that system could give rise to a dipolar anisotropy through the Compton-Getting effect \citep{cg}. For instance, for a CR distribution that is isotropic in the CMB rest frame, the resulting Compton-Getting dipole amplitude would be about 0.6\% \citep{cg2}. This amplitude depends on the relative velocity and on the CR spectral slope, but not directly on the particle charge.  The deflections of the extragalactic CRs caused by the Galactic magnetic field are expected to further reduce this amplitude, and also to generate higher harmonics, in a rigidity dependent way, so that the exact predictions are model dependent.  The Compton-Getting extragalactic contribution to the dipolar anisotropy is hence below the upper limits obtained.

More data, as well as analyses exploiting the discrimination between the different cosmic-ray mass components that will become feasible with the  upgrade  of the Pierre Auger Observatory currently being implemented \citep{augerprime}, will be crucial to understand in depth the origin of the cosmic rays at these energies and to learn how their anisotropies are produced. 

\acknowledgments

\section*{Acknowledgments}

\begin{sloppypar}
The successful installation, commissioning, and operation of the Pierre
Auger Observatory would not have been possible without the strong
commitment and effort from the technical and administrative staff in
Malarg\"ue. We are very grateful to the following agencies and
organizations for financial support:
\end{sloppypar}

\begin{sloppypar}
Argentina -- Comisi\'on Nacional de Energ\'\i{}a At\'omica; Agencia Nacional de
Promoci\'on Cient\'\i{}fica y Tecnol\'ogica (ANPCyT); Consejo Nacional de
Investigaciones Cient\'\i{}ficas y T\'ecnicas (CONICET); Gobierno de la
Provincia de Mendoza; Municipalidad de Malarg\"ue; NDM Holdings and Valle
Las Le\~nas; in gratitude for their continuing cooperation over land
access; Australia -- the Australian Research Council; Brazil -- Conselho
Nacional de Desenvolvimento Cient\'\i{}fico e Tecnol\'ogico (CNPq);
Financiadora de Estudos e Projetos (FINEP); Funda\c{c}\~ao de Amparo \`a
Pesquisa do Estado de Rio de Janeiro (FAPERJ); S\~ao Paulo Research
Foundation (FAPESP) Grants No.~2010/07359-6 and No.~1999/05404-3;
Minist\'erio da Ci\^encia, Tecnologia, Inova\c{c}\~oes e Comunica\c{c}\~oes (MCTIC);
Czech Republic -- Grant No.~MSMT CR LTT18004, LO1305, LM2015038 and
CZ.02.1.01/0.0/0.0/16{\textunderscore}013/0001402; France -- Centre de Calcul
IN2P3/CNRS; Centre National de la Recherche Scientifique (CNRS); Conseil
R\'egional Ile-de-France; D\'epartement Physique Nucl\'eaire et Corpusculaire
(PNC-IN2P3/CNRS); D\'epartement Sciences de l'Univers (SDU-INSU/CNRS);
Institut Lagrange de Paris (ILP) Grant No.~LABEX ANR-10-LABX-63 within
the Investissements d'Avenir Programme Grant No.~ANR-11-IDEX-0004-02;
Germany -- Bundesministerium f\"ur Bildung und Forschung (BMBF); Deutsche
Forschungsgemeinschaft (DFG); Finanzministerium Baden-W\"urttemberg;
Helmholtz Alliance for Astroparticle Physics (HAP);
Helmholtz-Gemeinschaft Deutscher Forschungszentren (HGF); Ministerium
f\"ur Innovation, Wissenschaft und Forschung des Landes
Nordrhein-Westfalen; Ministerium f\"ur Wissenschaft, Forschung und Kunst
des Landes Baden-W\"urttemberg; Italy -- Istituto Nazionale di Fisica
Nucleare (INFN); Istituto Nazionale di Astrofisica (INAF); Ministero
dell'Istruzione, dell'Universit\'a e della Ricerca (MIUR); CETEMPS Center
of Excellence; Ministero degli Affari Esteri (MAE); M\'exico -- Consejo
Nacional de Ciencia y Tecnolog\'\i{}a (CONACYT) No.~167733; Universidad
Nacional Aut\'onoma de M\'exico (UNAM); PAPIIT DGAPA-UNAM; The Netherlands
-- Ministry of Education, Culture and Science; Netherlands Organisation
for Scientific Research (NWO); Dutch national e-infrastructure with the
support of SURF Cooperative; Poland -Ministry of Science and Higher
Education, grant No.~DIR/WK/2018/11; National Science Centre, Grants
No.~2013/08/M/ST9/00322, No.~2016/23/B/ST9/01635 and No.~HARMONIA
5--2013/10/M/ST9/00062, UMO-2016/22/M/ST9/00198; Portugal -- Portuguese
national funds and FEDER funds within Programa Operacional Factores de
Competitividade through Funda\c{c}\~ao para a Ci\^encia e a Tecnologia
(COMPETE); Romania -- Romanian Ministry of Research and Innovation
CNCS/CCCDI-UESFISCDI, projects
PN-III-P1-1.2-PCCDI-2017-0839/19PCCDI/2018 and PN18090102 within PNCDI
III; Slovenia -- Slovenian Research Agency, grants P1-0031, P1-0385,
I0-0033, N1-0111; Spain -- Ministerio de Econom\'\i{}a, Industria y
Competitividad (FPA2017-85114-P and FPA2017-85197-P), Xunta de Galicia
(ED431C 2017/07), Junta de Andaluc\'\i{}a (SOMM17/6104/UGR), Feder Funds,
RENATA Red Nacional Tem\'atica de Astropart\'\i{}culas (FPA2015-68783-REDT) and
Mar\'\i{}a de Maeztu Unit of Excellence (MDM-2016-0692); USA -- Department of
Energy, Contracts No.~DE-AC02-07CH11359, No.~DE-FR02-04ER41300,
No.~DE-FG02-99ER41107 and No.~DE-SC0011689; National Science Foundation,
Grant No.~0450696; The Grainger Foundation; Marie Curie-IRSES/EPLANET;
European Particle Physics Latin American Network; and UNESCO.
\end{sloppypar}

\appendix

In Table~\ref{tab:sol-asid} we report the amplitudes and probabilities obtained with the SD1500 array at the solar and antisidereal frequencies, in all bins above 2~EeV for which the Rayleigh analysis was applied at the sidereal frequency. 
One can see that all these amplitudes are consistent with being fluctuations, showing then no signs of remaining systematic effects.
 We also report in Table~\ref{tab:dperew} the equatorial dipole amplitudes and phases obtained with the East-West method above 2~EeV, and compare them with the results for  the same datasets that were obtained with the Rayleigh method (reported in Table~\ref{tab:dper}).
The inferred equatorial dipole amplitudes turn out to be consistent, although the statistical uncertainty obtained with the East-West method is larger by a factor $\pi\langle \cos\delta\rangle/2\langle \sin\theta \rangle$ \citep{ew}.   Given that above full trigger efficiency one has that $\langle\sin\theta\rangle\simeq 0.58$ when considering $\theta<60^\circ$,  as we do for $E<4$~EeV, or $\langle\sin\theta\rangle\simeq 0.65$ when considering $\theta<80^\circ$,  as we do for $E\geq 4$~EeV, and that $\langle\cos\delta\rangle\simeq 0.78$ in both zenith ranges, the statistical uncertainties obtained in the East-West analysis are larger by a factor of about  2.1  than those obtained with the Rayleigh analysis for $\theta<60^\circ$, or by  a factor of about 1.9 for $\theta<80^\circ$, as can be seen in Table~\ref{tab:dperew}.

\begin{table}[h]
\centering
\begin{tabular}{c  c | c c |  c c }
    & & \multicolumn{2}{c |}{solar}  & \multicolumn{2}{c}{antisidereal}
 \\
    $ E$ [EeV] & $N$ &  $r$ [\%] & $P(\geq r)$  &  $r$ [\%] & $P(\geq r)$  
   \\
\hline
   2 - 4  & 283,074 & $0.6^{+0.3}_{-0.2}$ & 0.07 & $0.5^{+0.3}_{-0.2}$ & 0.20 
 \\
  4 - 8  & 88,325 & $0.8^{+0.5}_{-0.3}$ &0.24  & $0.5^{+0.5}_{-0.2}$ &0.59  
 \\
 8 - 16  & 27,271 & $0.6^{+1.1}_{-0.2}$ &0.79 & $0.5^{+1.1}_{-0.1}$ &0.83 
 \\ 
 16 - 32  & 7,664 & $1.1^{+2.0}_{-0.3}$ & 0.79 & $3.1^{+1.9}_{-1.1}$ & 0.16
 \\ 
    $\geq 32$  & 1,993 & $1.5^{+4.4}_{-0.1}$ & 0.90 & $1.3^{+4.6}_{-0.0}$ & 0.92 
 \\ 
 \hline
 $\geq 8$ & 36,928 & $0.3^{+1.1}_{-0.0}$ & 0.93  & $1.0^{+0.8}_{-0.4}$ & 0.39
 \\
 
\end{tabular}
\caption{Fourier amplitudes at the solar and antisidereal frequencies, and the probabilities to get larger values from statistical fluctuations of an isotropic distribution, for the different energy bins above 2~EeV.}
\label{tab:sol-asid}
\end{table}

\begin{table}[h]
\centering

\begin{tabular}{ c c | c c c c | c c c}
& & \multicolumn{4}{c |}{East-West (SD1500)} &  \multicolumn{3}{c}{Rayleigh (SD1500)} \\
    $E$ [EeV] & $N$ &  $d_\perp$ [\%] & $\sigma_{x,y}$ [\%] & $\alpha_d [^\circ]$ & $P(\ge d_\perp)$ &  $d_\perp$ [\%] &$\sigma_{x,y}$ [\%] &  $\alpha_d [^\circ]$\\
\hline
\rule{0pt}{3ex} 
 2 - 4  & 283,074 & $0.2^{+0.9}_{-0.2}$ & 0.72 & $-16 \pm 167$ & 0.94 & $0.5^{+0.4}_{-0.2}$ & 0.34 & $-11 \pm 55$ \\
 4 - 8  & 88,325 & $1.7^{+1.3}_{-0.7}$ &1.1 & $41 \pm 38$ & 0.33 &  $1.0^{+0.7}_{-0.4}$ & 0.61 & $69 \pm 46$\\
   8 - 16  & 27,271 & $6.4^{+2.3}_{-1.7}$ &2.1 & $147 \pm 18$ & $8.3\times 10^{-3}$ &  $5.6^{+1.2}_{-1.0}$ &1.1 & $ 97\pm 12$\\
   16 - 32  & 7,664 & $9.3^{+4.5}_{-3.0}$ & 3.9& $67 \pm 24$ & $5.8\times 10^{-2}$ &  $7.5^{+2.3}_{-1.8}$ & 2.1 &$ 80\pm 17$\\
    $\geq 32$  & 1,993 & $25^{+9}_{-6}$ & 7.6 & $151 \pm 17$ & $4.1\times 10^{-3}$ &  $13^{+5}_{-3}$ & 4.1 & $152 \pm 19$\\
\hline
\rule{0pt}{3ex} 
   $\geq 8$ & 36,928 & $6.6^{+2.0}_{-1.5}$ & 1.8& $132 \pm 15$ & $8.6\times 10^{-4}$ &  $6.0^{+1.0}_{-0.9}$ & 0.94 & $98 \pm 9$  \\
\end{tabular}
\caption{Equatorial dipole reconstruction above 2 EeV obtained using the East-West method.  Indicated are the number of events, amplitude of $d_\perp$, uncertainty $\sigma_{x,y}=\sigma/\langle\cos\delta\rangle$ of the components $d_x$ or $d_y$, right ascension phase $\alpha_d$  and probability $P(\geq d_\perp)$ to get a larger amplitude from fluctuations of an isotropic distribution. For comparison we also include in the last two columns the values of $d_\perp$ and $\alpha_d$ that were obtained in the Rayleigh analysis (reported in Table~\ref{tab:dper}).}
\label{tab:dperew}
\end{table}

\eject

\end{document}